# Multi-frequency phase retrieval from noisy data

Vladimir Katkovnik and Karen Egiazarian
Laboratory of Signal Processing, Tampere University of Technology, P.O. Box 553
FI-33101 Tampere, Finland
e-mails: vladimir.katkovnik@tut.fi, karen.egiazarian@tut.fi

*Abstract*—The phase retrieval from multi-frequency intensity (power) observations is considered. The object to be reconstructed is complex-valued. A novel algorithm is presented that accomplishes both the object phase (absolute phase) retrieval and denoising for Poissonian and Gaussian measurements. The algorithm is derived from the maximum likelihood formulation with Block Matching 3D (BM3D) sparsity priors. These priors result in two filtering: one is in the complex domain for complex-valued multi-frequency object images and another one in the real domain for the object phase. The algorithm is iterative with alternating projections between the object and measurement variables. The simulation experiments are produced for Fourier transform image formation and random phase modulations of the object, then the observations are random object diffraction patterns. The results demonstrate the success of the algorithm for reconstruction of the complex phase objects with the high-accuracy performance even for very noisy data.

## I. INTRODUCTION

Phase retrieval is a problem concerning reconstruction of the phase from intensity (power) measurements of complex-valued variables. It is inherently ill-posed due to the lack of the phase in observations. The fields of application are varying from physics and engineering to medicine and biology, e.g. [1]-[3].

We consider the problem of $2D$ imaging from the multi-frequency observations. The multi-frequency complex-valued object model is of the form

$$u_{o,\lambda} = b_{o,\lambda} \exp(j\mu_\lambda \varphi), \ \lambda \in \Lambda, \quad (1)$$

where $u_{o,\lambda} \in \mathbb{C}^{N \times N}$, $\varphi$ is the object absolute phase in radians, $\mu_\lambda > 0$ are dimensionless relative frequencies and $\Lambda = [\lambda_o, \lambda_1, ..., \lambda_{n_\lambda - 1}]$ is a set of periods (wavelengths) of the multi-frequency observations. In what follows, the amplitude, phase and other variables are functions of the argument $x$ given on a regular $2D$ grid, $X \subset \mathbb{Z}^2$.

In the model (1), $\mu_\lambda$ is a frequency depending scale parameter of the object phase which establishes a link between the absolute phase $\varphi$ and the wrapped phase $\psi_\lambda$ of $u_{o,\lambda}$ which can be measured at the $\lambda$-channel. The wrapped phase is related with the true absolute phase $\varphi$ as $\mu_\lambda \varphi = \psi_\lambda + 2\pi k_\lambda$, where $k_\lambda$ is an integer, $\psi_\lambda \in [-\pi, \pi)$. The link between the absolute and wrapped phase is installed as:

$$\psi_\lambda = \mathcal{W}(\mu_\lambda \varphi) \equiv \mathrm{mod}(\mu_\lambda \varphi + \pi, 2\pi) - \pi.$$

where $\mathcal{W}(\cdot)$ is the wrapping operator, which decomposes the absolute phase $\mu_\lambda \varphi$ into two parts: the fractional part $\psi_\lambda \in [\pi, -\pi)$ and the integer part defined as $2\pi k_\lambda$.

The frequencies, i.e. the phase factors $\mu_\lambda$ are given but the phase $\varphi$ and the amplitude $b_{o,\lambda}$ should be retrieved from the intensity observations, which are for the noiseless $y_{s,\lambda}$ and noisy $z_{s,\lambda}$ observations have the form:

$$u_{s,\lambda} = \mathcal{P}_{s,\lambda} u_o, \ y_{s,\lambda} = |u_{s,\lambda}|^2, \quad (2)$$
$$z_{s,\lambda} = \mathcal{G}\{|u_{s,\lambda}|^2\}, \lambda \in \Lambda, \ s = 1, ..., S, \quad (3)$$

where $S$ is a number of experiments.

The image formation operator $\mathcal{P}_{s,\lambda}$, usually depending on $\lambda$, is also known. The observations are given for each $\lambda \in \Lambda$. If the phases $\mu_\lambda \varphi$ are interferometric, i.e. $\mu_\lambda \varphi \in [\pi, -\pi)$, then the problem becomes trivial as the estimates of $\mu_\lambda \varphi$ and $b_{o,\lambda}$ can be found separately for each $\lambda$ by one of the phase retrieval algorithms applicable for complex domain imaging. The problem becomes much more challenging, when the phases $\mu_\lambda \varphi$ are non-interferometric, $\mu_\lambda \varphi \notin [\pi, -\pi)$ because the data processing produced independently for each $\lambda$ gives estimates only for the wrapped phases $\psi_\lambda$. Aggregation of these estimates in order to reconstruct the absolute phase $\varphi$ is a crucial issue of the multi-frequency phase retrieval problem.

*A. Related works*

For a single frequency, $\mu_\lambda = 1$ and $b_{o,\lambda} = b_o$ in (1), the multi-frequency problem becomes the conventional phase retrieval: to reconstruct the object $u_o \in \mathbb{C}^{N \times N}$ from noiseless $y_s$ or noisy $z_s$ intensity observations. A flow of publications old and recent concern various formulations and algorithms for solution of this conventional setup starting from the classic Gerchberg-Saxton (GS) style algorithms (e.g. [4], [5]) based on alternating projections between the complex-valued $u_o$ and $u_s$. In this iterations the amplitudes in $u_s$ are tuned to the given observations. The back projections are modified accordingly to the prior information on the object such as a support size and shape, amplitude of $u_o$, etc. The review and analysis of the GS algorithms as well as further developments can be seen in [6]. The variational formulations of the phase retrieval have a stronger mathematical background and go to solving optimization tasks [7], [8]. The maximum likelihood approaches using the transform domain phase/amplitude sparsity are proved to be effective for noisy data. It is a base of the Sparse Phase Amplitude Retrieval (SPAR) algorithms [9]-[12].

The multi-frequency phase retrieval is a much less studied problem. The lower synthetic frequencies (beat-frequencies) appear in the differences of the wrapped phases $\Delta_{\lambda,\lambda'} = \mathcal{W}(\mu_\lambda \varphi) - \mathcal{W}(\mu_{\lambda'} \varphi) = \mathcal{W}((\mu_\lambda - \mu_{\lambda'})\varphi)$. The problem can be solved in the straightforward manner provided that

$\mu_\lambda - \mu_{\lambda'}$ is small and $(\mu_\lambda - \mu_{\lambda'})\varphi \in [\pi, -\pi)$, then $\varphi$ can be easily calculated from $\Delta_{\lambda, \lambda'}$. However, the noise level in these synthetic frequency estimates grows proportionally to $1/(\mu_\lambda - \mu_{\lambda'})$ [13]. It is a drawback of this approach. The various forms of the synthetic frequency techniques are proposed for multi-frequency scenarios (e.g. [14], [15]). The $2D$ phase unwrapping algorithms, i.e. absolute phase retrieval with simultaneous processing of multi-frequency noisy complex-exponents, have been developed in terms of the maximum likelihood formulation [16], [17]. The Chinese Remainder Theorem provides a wide class of the methods with a good theoretical background [18]. Reformulation of these approaches to noisy data and for robust estimation leads to the techniques similar to the maximum likelihood methods [19], [20].

In this paper we propose the approach and derive the algorithm which can be treated as development of the SPAR algorithm [11]-[12] and the maximum likelihood absolute phase reconstruction from multi-frequency observations [17].

*B. Contribution of this paper*

We present a novel multi-frequency phase retrieval algorithm that accomplishes both phase unwrapping and denoising from intensity observations. The noise models account for Poissonian and Gaussian distributions. The proposed algorithm is iterative with alternating projections between object and measurement variables. The algorithm is derived from the maximum likelihood formulation with the Block Matching 3D (BM3D) sparsity priors. These priors result in the two types of filtering: one is in the complex domain for multi-frequency object images and another one is in the real domain for the object absolute phase (object shape). One of the paramount issues of the algorithm is the absolute phase reconstruction by aggregation of the object multi-frequency estimates. This aggregation is achieved by the least square solution including compensation of the invariant phase-shifts in the $\lambda$-channel object estimates.

The simulation experiments are produced for Fourier transform image formation with the random phase modulations of the object. The observations are noisy random object diffraction patterns. The results demonstrate the success of the algorithm for reconstruction of the high complexity phase objects with the precise performance even for very noisy observations.

## II. ALGORITHM DEVELOPMENT

*A. Problem formulation*

Reconstruction of $u_o \in \mathbb{C}^{N \times N}$ from noisy $\{z_{s,\lambda}\}$ is rather challenging mainly due the periodic nature of the likelihood function with respect to the phase $\varphi$ and the non-linearity of the observation model. Provided a stochastic noise model with independent samples, the maximum likelihood leads to the basic criterion $\mathcal{L}_0 = \sum_{\lambda \in \Lambda} \sum_{s=1}^{S} \sum_x l(z_{s,\lambda}(x), |u_{s,\lambda}(x)|^2)$, where $l(z, |u|^2)$ denotes the minus log-likelihood of a candidate solution for $u_o$ given through the observed true intensities $|u|^2$ and noisy outcome $z$. For the Poissonian and Gaussian distributions we have, respectively, $l(z, |u|^2) = |u|^2 \chi -$ $z \log(|u|^2 \chi)$, $\chi > 0$ is the noise scaling parameter [11] and $l(z, |u|^2) = \frac{1}{2\sigma^2}||u|^2 - z|^2$, $\sigma^2$ stands for the noise variance.

We introduce the following criterion including the image formation model $\mathcal{P}_{s,\lambda}$ for $u_{s,\lambda}$ and the complex-valued exponent modeling of the multi-frequency object $u_{o,\lambda}$:

$$\mathcal{L}(u_{s,\lambda}, u_{o,\lambda}, b_o, \varphi, \delta_\lambda) = \sum_{\lambda,s,x} l(z_{s,\lambda}(x), |u_{s,\lambda}(x)|^2) +$$
$$\frac{1}{\gamma_1} \sum_{\lambda,s} ||u_{s,\lambda} - \mathcal{P}_{s,\lambda} u_{o,\lambda}||_2^2 + \quad (4)$$
$$\frac{1}{\gamma_2} \sum_\lambda ||b_{o,\lambda} \exp(j(\mu_\lambda \cdot \varphi + \delta_\lambda)) - u_{o,\lambda}||_2^2,$$

where $\gamma_1, \gamma_2 > 0$ are parameters, and $||\cdot||_2^2$ is the Hadamard norm. The maximum likelihood in (4) is penalized by the quadratic residual function for correspondence of $u_{s,\lambda}$ to $u_{o,\lambda}$.

The estimates of $\mu_\lambda \varphi(x)$ in (4) can be biased due to the fact that in the phase retrieval the solution is defined only within an invariant additive phase shift. We model this biasedness by the parameters $\delta_\lambda$ invariant with respect to $x$ and additive to the phases $\mu_\lambda \hat{\varphi}$.

We say that the object complex exponents $u_{o,\lambda}$ are in-phase (synchronized) if $\delta_\lambda = 0$, for all $\lambda \in \Lambda$, and out-of-phase otherwise. The phase shifts $\delta_\lambda$ should be compensated for a proper estimation of the absolute phase $\varphi$.

The criterion $\mathcal{L}$ is minimized with respect to $u_{s,\lambda}, u_{o,\lambda}, b_o, \varphi$ as well as with respect to the phase-shifts $\delta_\lambda$.

1) Minimization with respect to $u_{s,\lambda}$ concerns the first two summands in $\mathcal{L}$. The problem is additive and can be obtained separately for each $x$, $s$ and $\lambda$. The corresponding analytical solutions derived for both the Poissonian and Gaussian distributions can be seen in [11]. These solutions optimal in the maximum likelihood sense define $u_{s,\lambda}$ as functions of noisy observation $z_{s,\lambda}$ and projections $\mathcal{P}_s u_{o,\lambda}$ of $u_{o,\lambda}$ on the sensor.

2) Minimization with respect to $u_{o,\lambda}$ goes to the last two summands of the criterion. It is the quadratic problem. If $\mathcal{P}_{s,\lambda}$ are orthonormal such that $\sum_s \mathcal{P}^*_{s,\lambda} \mathcal{P}_{s,\lambda} = I$, where $I$ is the identity operator and $\mathcal{P}^*_{s,\lambda}$ is Hermitian adjoint for $\mathcal{P}_{s,\lambda}$, the solution is of the form

$$\hat{u}_{o,\lambda} = \frac{\sum_s \mathcal{P}^*_{s,\lambda} u_{s,\lambda} + \gamma_1/\gamma_2 b_{o,\lambda} \exp(j\mu_\lambda \varphi)}{1 + \gamma_1/\gamma_2}. \quad (5)$$

3) Minimization on $b_{o,\lambda}$, $\varphi$ and $\delta_\lambda$ concerning the last summand in $\mathcal{L}$ is a non-linear least square fitting of the frequency dependent $u_{o,\lambda}$. Minimization on $\varphi$ is an absolute phase estimation, i.e. the phase unwrapping problem. The solution is composed from the following two successive stages $A$ and $B$.

(A) *Phase synchronization*. Let $\lambda' \in \Lambda$ be a reference channel. Define the estimates for the phase shift $\delta_\lambda$ for (4) in the following way

$$\hat{\delta}_\lambda = \mathcal{W}(\hat{\delta}_{\lambda'} \cdot \mu_\lambda / \mu_{\lambda'} + \delta_{\lambda, \lambda'} \mu_\lambda), \quad (6)$$

where $\hat{\delta}_{\lambda'}$ is a hypothetical value of unknown $\delta_{\lambda'}$ and $\delta_{\lambda,\lambda'}$ is a shift between the weighted wrapped phases of the reference and $\lambda$-channels

$$\delta_{\lambda,\lambda'} = median_x(\mathcal{W}(\psi_{o,\lambda}(x)/\mu_\lambda - \psi_{o,\lambda'}(x)/\mu_{\lambda'})). \tag{7}$$

One of the goals is to reduce the number of unknown phase shifts $\delta_\lambda$, $\lambda \in \Lambda$, to a single $\delta_{\lambda'}$, i.e. to the phase-shift in the reference channel.

It was tested by Monte-Carlo modeling that $\delta_{\lambda,\lambda'}$ is a good estimate for $\delta_\lambda/\mu_\lambda - \delta_{\lambda'}/\mu_{\lambda'}$, then $\hat{\delta}_\lambda = \delta_\lambda + (\hat{\delta}_{\lambda'} - \delta_{\lambda'})\mu_\lambda/\mu_{\lambda'}$. Inserting this $\hat{\delta}_\lambda$ in (4) instead of $\delta_\lambda$ we arrive to

$$\mathcal{L}_1(b_{o,\lambda}, \varphi, \delta_\lambda) = \tag{8}$$
$$\sum_\lambda ||b_{o,\lambda}\exp(j(\mu_\lambda(\varphi + \Delta\varphi)) -$$
$$|u_{o,\lambda}|\exp(j(\psi_{o,\lambda} - \delta_\lambda))||_2^2, \ \Delta\varphi = (\hat{\delta}_{\lambda'} - \delta_{\lambda'})/\mu_{\lambda'}.$$

The accurate synchronization of the exponents $\exp(j(\mu_\lambda(\varphi + \Delta\varphi))$ is achieved in (8) as a result of compensation by $\hat{\delta}_\lambda$ the phase shifts existing in the wrapped phases $\psi_{o,\lambda}$. If $\Delta\varphi \neq 0$ the phase $\varphi$ is estimated within an invariant $\Delta\varphi$. It is not impose any restrictions as by default the phase $\varphi$ in the phase-less measurements of the phase retrieval can be estimated within an arbitrary invariant shift only.

(B) *Absolute phase retrieval (APR)*. In order to simplify the presentation assume that the complex exponents of $u_{o,\lambda}$ are perfectly in phase, i.e. $\Delta\varphi = 0$, and $b_{o,\lambda} = |u_{o,\lambda}|$, then the absolute phase is reconstructed as

$$\hat{\varphi} = \tag{9}$$
$$\arg\min_\varphi \sum_\lambda |u_{o,\lambda}|^2[1 - \cos(\mu_\lambda \cdot \varphi - \hat{\psi}_{o,\lambda})],$$

where $\hat{\psi}_{o,\lambda} - \hat{\delta}_\lambda$ is denoted as $\hat{\psi}_{o,\lambda}$, i.e. the $\delta_\lambda$ unknown in (8) are replaced by their estimates $\hat{\delta}_\lambda$.

For calculation of $\hat{\varphi}$ we use the approach proposed in [17]. Due to the periodicity of the cosines in (9) $\mu_\lambda \cdot \varphi = \hat{\psi}_{o,\lambda} + 2\pi k_\lambda$, where $k_\lambda$ are integers. By summation over $\lambda$ we obtain $\varphi = \dfrac{\sum_\lambda(\hat{\psi}_{o,\lambda} + 2\pi k_\lambda)}{\sum_\lambda \mu_\lambda}$ and substituting this $\varphi$ into (9) we arrive to

$$\hat{\varphi} = \arg\min_{k \in [0, Q)} \sum_\lambda |u_{o,\lambda}|^2[1 - \tag{10}$$
$$\cos(\frac{\mu_\lambda(\sum_\lambda \hat{\psi}_{o,\lambda} + 2\pi k)}{\sum_\lambda \mu_\lambda} - \hat{\psi}_{o,\lambda})], \ k = \sum_\lambda k_\lambda,$$
$$\hat{b}_{o,\lambda} = |u_{o,\lambda}|. \tag{11}$$

It shows that the multivariable optimization on $k_\lambda$ is reduced to the scalar optimization on the integer $k$. If $\mu_\lambda = p_\lambda/q_\lambda$, where $(p_\lambda, q_\lambda)$ are coprime integer, then the criterion in (10) is a periodic function of $k$ with the synthetic period $Q$ equal to the nominator of $\sum_\lambda \mu_\lambda$. It makes clear why we restrict the optimization in (10) to the interval $[0, Q)$. We obtain from the above formula for $\varphi$ that the maximal range of the absolute $\varphi$ which can be reconstructed in this approach depends on the spectrum of the multiple frequencies and restricted by $2\pi(n_\Lambda + Q)/\sum_\lambda \mu_\lambda$, where $n_\Lambda$ is a number of the frequencies in $\Lambda$. Rational approximations are used for $\mu_\lambda$ in these calculations if $\mu_\lambda$ are not rational.

### B. Algorithm's implementation

Using the above solutions the iterative algorithm is developed of the structure shown in Table I. The initialization by the complex-valued $u_{o,\lambda}^1$ is obtained from the observations $\{z_{s,\lambda}\}$ by the SPAR algorithm [11] separately for each $\lambda$. The main iterations start from the forward propagation (Step 1) and follows by the amplitude update for $u_{s,\lambda}^t$ at Step 2. The operator $\Phi_1$ for this update is obtained by minimization of $\mathcal{L}$ on $u_{s,\lambda}$. It can be seen in [11] for Poissonian and Gaussian noise models. The back propagation in Step 3, the operator $\Phi_2$, is defined by Eq.(5). The absolute phase reconstruction from the wrapped phases of $u_{o,\lambda}^{t+1}$ is produced in Step 4 by the $ARP$ algorithm as defined by the solutions (10)-(11). The obtained amplitude and phase update $u_{o,\lambda}^{t+1}$ at Step 5. The number of iteration is fixed in this implementation of the algorithm. The steps 3 and 4 are completed by the BM3D filtering. In Step 3 it is the filtering of complex-valued $u_{o,\lambda}^{t+1/2}$ produced separately for the wrapped phase and amplitude of $u_{o,\lambda}^{t+1/2}$. In Step 4 this filtering is applied to the absolute phase $\hat{\varphi}^{t+1/2}$. These BM3D filters are derived for the considered phase retrieval problem from the group-wise sparsity priors for the filtered variables. This technique is based on the Nash equilibrium formulation for the phase retrieval instead of the more conventional constrained optimization with a single criterion function as it is, for instance, in (4). We do not show this derivation as it is quite similar to the derivation presented in [11].

TABLE I
MULTI-FREQUENCY ABSOLUTE PHASE RETRIEVAL (MF-APR) ALGORITHM

| | |
|---|---|
| | **Input**: $\{z_{s,\lambda}\}$, $s = 1, ..., S$, $\lambda \in \Lambda$, |
| | **Initialization**: $u_{o,\lambda}^1$, $\lambda \in \Lambda$ |
| | **Main iterations:** $t = 1, 2, ..., T$ |
| 1. | **Forward propagation**: |
| | $u_{s,\lambda}^{t+1/2} = \mathcal{P}_{s,\lambda} u_{o,\lambda}^t$, $s = 1, ..., S$, $\lambda \in \Lambda$; |
| 2. | **Noise suppression and update of $u_{s,\lambda}^t$**: |
| | $u_{s,\lambda}^t = \Phi_1(u_{s,\lambda}^{t+1/2}, z_{s,\lambda})$; |
| 3. | **Backward propagation and filtering:** |
| | $u_{o,\lambda}^{t+1/2} = \Phi_2(u_s^t)$, $u_{o,\lambda}^t = BM3D(u_{o,\lambda}^{t+1/2})$; |
| 4. | **Absolute phase retrieval and filtering:** |
| | $\{\varphi^{t+1/2}, b_{o,\lambda}^{t+1}\} = \mathcal{APR}(u_{o,\lambda}^t)$, |
| | $\varphi^{t+1} = BM3D(\varphi^{t+1/2})$; |
| 5. | **Object update:** |
| | $u_{o,\lambda}^{t+1} = b_{o,\lambda}^{t+1}\exp(j\varphi^{t+1}\mu_\lambda)$, $\lambda \in \Lambda$; |
| | **Output**: $\varphi^{T+1}$, $b_{o,\lambda}^{T+1}$, $u_{o,\lambda}^{T+1}$. |

In our experiments the parameters of the algorithm are fixed for all tests. The parameters defining the iterations of the algorithm are as follows: $\gamma_1 = 1/\chi$, where $\chi$ is the parameter of the Poissonian distribution, $\gamma_1/\gamma_2 = 0.2$. The parameters of BM3D filters can be seen in [11]. The MATLAB demo-codes of the algorithm are publicly available [1].

## III. NUMERICAL EXPERIMENTS

In numerical experiments we model a lensless optical system where a thin flat transparent phase object is illuminated by monochromatic three color RGB (red-green-blue) coherent light beams from lasers or LEDs. The wavefronts just behind the object are $u_{o,\lambda} = b_{o,\lambda} \exp(j2\pi \cdot h \cdot (n_\lambda - 1)/\lambda)$, where $h \geq 0$ is an object thickness (height, shape, profile), $\lambda$ is the wavelength, and $n_\lambda$ is a refractive index of the object material. We recalculate the wavelengths including the effect of the refractive index as $\lambda \rightarrow \lambda/(n_\lambda - 1) \in \Lambda$, $\Lambda = [500, 600, 700] \times 10^{-9}$ m. Then, the formula $2\pi \cdot h/\lambda$ defines the phase delay in propagation of the coherent light-beam through the object. In notation used in (1), the relative frequencies are $\mu_\lambda = \lambda_{ref}/\lambda$, where $\lambda_{ref}$ is a reference wavelength, and $\varphi = 2\pi h/\lambda_{ref}$ is the reference absolute phase in radians.

The propagation of the wavefronts to the sensor is given by the Fourier transform [21]. The intensities of the light beam impinging on the sensor parallel to the object plane are calculated as $z_{o,\lambda} = \mathcal{G}\{|\mathcal{F}\{M_s \circ u_{o,\lambda}\}|^2\}$, $s = 1, ..., S$, $\lambda \in \Lambda$. Here $M_s$ are the modulation phase-masks inserted next to the object, $'\circ'$ stands for the pixel-wise multiplication of the object and phase-mask transition functions. The phase-masks $M_s$ enable strong diffraction of the wavefield propagation and are introduced in order to achieve the phase diversity sufficient for improved reconstruction of the complex-valued object. Similar to [22], we use the four phase values $[0, \pi/2, \pi, 3/2\pi]$ prescribed to the pixels of the modulation phase-masks randomly.

The accuracy of the object reconstruction is characterized by the Relative Root-Mean-Square Error ($RRMSE$) criteria calculated as $RMSE$ divided by the root of the mean square power of the signal. In these criteria the biasedness of the phase estimate is corrected by the mean value of the error between the estimate and the true value.

We show the results obtained for Poissonian observations and very noisy data. The noisiness of observations is characterized by Signal-to-Noise Ratio (SNR) $SNR = 10\log_{10}(\chi^2 \sum_{s,\lambda} ||u_{s,\lambda}||_2^2 / \sum_{s,\lambda} ||\chi \cdot |u_{s,\lambda}|^2 - z_{s,\lambda}||_2^2)$ $dB$ and by the mean number of photons per sensor pixel, $N_{photon}$.

The illustrating results are presented for two phase objects with the invariant amplitude equal to 1: Gaussian ($100 \times 100$) and U.S. Air Force (USAF) resolution target ($612 \times 612$).

For the given $\Lambda$ and the reference wavelength $\lambda_{ref} = \lambda_1$ we have $\mu_1 = 1$, $\mu_2 = 0.0833$, $\mu_3 = 0.7143$. The rational approximations for these $\mu_\lambda$ are $\mu_1 = 1$, $\mu_2 = 5/6$, $\mu_3 = 5/7$ and the sum $\sum_\lambda \mu_\lambda = 107/42$. Thus, $Q = 107$ and the upper bound for the range of $\varphi$ is calculated as $2\pi(n_\Lambda +$

[1] http://www.cs.tut.fi/sgn/imaging/sparse/

$Q)/\sum_\lambda \mu_\lambda \simeq 272\ rads$. The number of experiments for each wavelength is $S = 6$.

The peak-value of the Gaussian phase for the reference wavelength is $\varphi = 242$ radians, what is close to the upper limit $272\ rads$. To make the problem more difficult we consider a very noisy scenario $SNR = 12.5$ dB and $N_{photon} = 7$.

The achieved phase reconstructions are shown in Fig.1 as $3D/2D$ images. The proposed algorithm produced a nearly perfect reconstruction with $RRMSE = 0.00529$ while counterpart reconstructions are completely failed for all wavelengths. These latter absolute phase reconstructions are obtained by the SPAR algorithm as the wrapped phase estimates unwrapped further by the PUMA algorithm [23] for each wavelength separately. The true wrapped phase pattern in this test is so complex and so irregular (see. Fig.2) that it is just impossible to unwrap it by any modern 2D algorithms. It is not surprising that PUMA, which is one of the best algorithms in the field, failed. The proposed MF-APR algorithm is able to resolve the problem even for the noisy data.

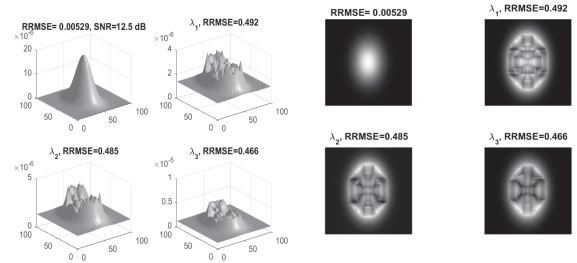

Fig. 1. Gaussian absolute phase reconstructions, 3D/2D images: left-top, by the proposed MF-APR algorithms with the best RRMSE value, others are absolute phases obtained from the wrapped phases given by the SPAR algorithm and unwrapped by the PUMA algorithm.

The phase reconstructions for the USAF test-object are shown in Fig.3 as $3D/2D$ phase images for noisy observations with $SNR = 8.74$ dB and $N_{photon} = 3.74$. The proposed algorithm demonstrates a quite accurate reconstruction with $RRMSE = 0.063$. The counterparts produced as above for each wavelength separately are failed.

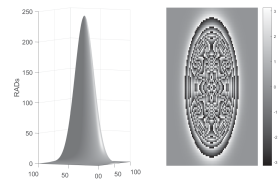

Fig. 2. The $3D$ true absolute phase of the Gaussian phase object, and the corresponding $2D$ wrapped phase, the wavelength $\lambda_{ref} = \lambda_1$.

The peak-value of the phase for this test with the reference wavelength $\lambda' = \lambda_1$ is $\varphi = 30$ rads. It is not so high as is for the Gaussian phase image but this test images is discontinuous binary with noisy observations what defines the complexity of the problem at hand.

The synthetic wavelength approach discussed in Section I-A is not effective for the considered tests as the wrapped phase differences are not small enough in order to $(\mu_\lambda - \mu_{\lambda'})\varphi \in [\pi, -\pi]$. Thus, it cannot be a valuable alternative to the proposed algorithm. For our experiments we use MATLAB R2015b and the computer with the processor Intel(R) Core(TM) i7-4800MQ@ 2.7 GHz. The complexity of the algorithm is characterized by the computational time. For $100 \times 100$ and $612 \times 612$ images, the computation time is 3.3 and 14.6 sec. per iteration. The APR algorithm takes, respectively, 0.2 sec. and 8.5 sec. of this time.

## IV. Conclusion

The multi-frequency absolute phase retrieval from intensity observations is considered. This paper introduces a variational approach to object phase and amplitude reconstruction from noisy intensity observations. The maximum likelihood criterion used in the developed optimization approach defines the intention to reach statistically optimal solutions. The phase retrieval is an ill-possed inverse problem, where the observation noise is amplified and transferred to estimates of phase and amplitude. The sparsity developed for modeling of varying variables is one of the key instruments for regularization of this inverse problem. The block-wise sparsity implemented by BM3D filters applied to amplitude/phase of complex-valued variables and the real-valued object absolute phase. One of the key original elements of the algorithm is the weighted least square solution for aggregation of the multi-frequency estimates developed for both the absolute phase reconstruction from the observed multi-frequency wrapped phases and the phase-shifts compensation. The effectiveness of the developed algorithm is demonstrated by simulation experiments for the coded diffraction pattern scenario and very noisy Poissonian observations. The MATLAB demo-codes of the algorithm are publicly available.

## V. Acknowledgment

This work is supported by Academy of Finland, project no. 287150, 2015-2019.

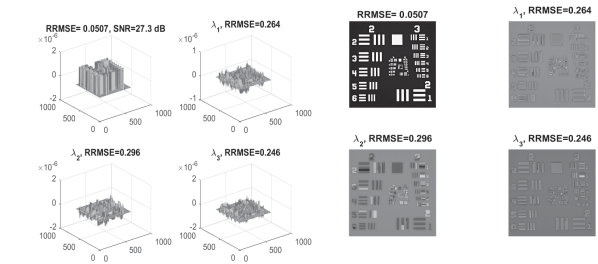

Fig. 3. USAF phase reconstructions, 3D/2D images: left-top, by the proposed MF-APR algorithms with the best RRMSE value, others are absolute phases obtained from the wrapped phases given by the SPAR algorithm and unwrapped by the PUMA algorithm.